\documentclass[twoside,journey]{IEEEtran}

\usepackage{makecell}
\usepackage{hyperref}
\usepackage{array}
\usepackage{caption2}
\usepackage{graphicx,amssymb,amsmath}
\usepackage{multicol}
\usepackage[noadjust]{cite}
\usepackage{setspace}
\usepackage{subfigure}
\usepackage{graphicx}
\usepackage{float}
\usepackage{url}
\usepackage{stfloats}
\usepackage{amsthm,pifont}
\usepackage{flushend}
\usepackage{cases,subeqnarray}
\usepackage{bm,multirow,bigstrut}
\usepackage{amsmath, amsthm, amssymb}
\usepackage{textcomp}
\usepackage{latexsym,bm}
\usepackage{booktabs}
\usepackage{xcolor}
\usepackage{mathtools}
\usepackage{dsfont}
\usepackage{extarrows}
\usepackage{epsfig}
\usepackage{epsfig}
\usepackage{epstopdf}
\usepackage[noend]{algpseudocode}
\usepackage{algorithmicx,algorithm}
\usepackage{calc}
\theoremstyle{plain}

\theoremstyle{plain}

\usepackage{amsmath}

\IEEEoverridecommandlockouts

\begin{document}
\title{Vision-based Semantic Communications for Metaverse Services: A Contest Theoretic Approach}
\author{Guangyuan Liu, Hongyang Du, Dusit Niyato,~\IEEEmembership{Fellow,~IEEE}, Jiawen~Kang, Zehui~Xiong, and Boon Hee Soong
\thanks{G.~Liu, H.~Du, and D. Niyato are with the School of Computer Science and Engineering, Nanyang Technological University, Singapore (e-mail: liug0022@e.ntu.edu.sg, hongyang001@e.ntu.edu.sg, dniyato@ntu.edu.sg)}
\thanks{J. Kang is with the School of Automation, Guangdong University of Technology, China. (e-mail: kavinkang@gdut.edu.cn)}
\thanks{Z. Xiong is with the Pillar of Information Systems Technology and Design, Singapore University of Technology and Design, Singapore (e-mail:
zehui\_xiong@sutd.edu.sg)}
\thanks{B. H. Soong is with the Department of Electrical and Electronic Engineering, Nanyang Technological University, Singapore (e-mail: ebhsoong@ntu.edu.sg)}

}
\maketitle
\vspace{-1cm}
\begin{abstract}
The popularity of Metaverse as an entertainment, social, and work platform has led to a great need for seamless avatar integration in the virtual world. In Metaverse, avatars must be updated and rendered to reflect users' behaviour. Achieving real-time synchronization between the virtual bilocation and the user is complex, placing high demands on the Metaverse Service Provider (MSP)'s rendering resource allocation scheme. To tackle this issue, we propose a semantic communication framework that leverages contest theory to model the interactions between users and MSPs and determine optimal resource allocation for each user. To reduce the consumption of network resources in wireless transmission, we use the semantic communication technique to reduce the amount of data to be transmitted. Under our simulation settings, the encoded semantic data only contains 51 bytes of skeleton coordinates instead of the image size of 8.243 megabytes. Moreover, we implement Deep Q-Network to optimize reward settings for maximum performance and efficient resource allocation. With the optimal reward setting, users are incentivized to select their respective suitable uploading frequency, reducing down-sampling loss due to rendering resource constraints by 66.076\% compared with the traditional average distribution method. The framework provides a novel solution to resource allocation for avatar association in VR environments, ensuring a smooth and immersive experience for all users.
\end{abstract}


\section{Introduction}
Metaverse is a shared and persistent three-dimensional (3D) virtual reality (VR) space that integrates multiple technologies, such as sensing, communication, and computing while integrating virtuality and reality. Based on the initial concept proposed in Neal Stephenson's 1992 science fiction novel \textit{Snow Crash}, there has been substantial growth in the popularity of Metaverse as an entertainment, social, and working platform~\cite{9806418}. Undoubtedly, such popularity has led to the need for seamless avatar integration in the virtual world. As the first step, avatars need to be updated and rendered to represent users in Metaverse in a way that accurately reflects their behaviour. Recently, with the advancement of Artificial Intelligence (AI), several studies~\cite{avatar1,avatar2} have deployed Human pose estimation (HPE) as a tool to acquire users' behaviour information for avatar association purposes.

Even though the deployment of HPE for avatar association has been researched, the allocation of server resources for users' avatar synchronization in such scenarios has yet to be well discussed~\cite{yang2022metafi}. Here, resource allocation refers to distributing computational resources, such as rendering power and memory . In HPE for avatar association, it is pertinent to consider resource allocation, as the MSPs must allocate sufficient resources to accurately update the avatars of all users in real-time. This becomes increasingly relevant in virtual environments where high-fidelity, real-time avatar action is crucial for providing an immersive experience. The MSPs must ensure that computational resources are distributed fairly among users and that resource constraints negatively impact no single user's experience. Additionally, the MSP should allocate resources dynamically based on changes in the number of users and the complexity of their movements. In this way, MSP can provide a accommodation between different services as well as different users with a better Quality of Service (QoS).

In response to this need, we propose a contest theory-based semantic communication framework. This framework leverages the principles of contest theory, i.e., a non-cooperative game theory that studies decision-making in situations where multiple individuals or users have conflicting interests and do not cooperate to achieve a common goal. In the context of HPE information for avatar association, the conflicting interests are each user's updating demands and the server's limited rendering resources. The contributions of this paper can be summarised as 
\begin{itemize}
    \item We deploy HPE to perform semantic encoding of pose information. This encoding optimizes the amount of data transmitted over the network by reducing the amount of data transmitted. The test results that demonstrate the encoding schemes' effectiveness are also presented.
    \item We propose a framework that uses a game-theoretic approach to model the interactions between the users and the MSPs and determines an optimal resource allocation for each user. The framework considers factors such as the rendering demands, i.e., each user's movement, the server's available resources, and the desired QoS.  By implementing this contest theory-based semantic communication framework, we aim to provide a novel solution to the challenge of resource allocation in deploying HPE for avatar association and ensure that the VR experience is smooth and immersive for all users.
    \item We deployed Deep Q-Network (DQN) to optimize award settings for minimized down-sampling loss induced by limited resources. The results show the effectiveness of DQN for reward optimization and its ability to minimize fluctuations caused by discrete action space and the total down-sampling loss. The framework provides a novel solution to resource allocation for avatar association in VR environments. 
\end{itemize}

\section{Related works}
\subsection{Dynamic resource allocation for MSP}
The growing Metaverse market has increased in virtual services, including socializing, entertainment and education~\cite{Xu_Ng_Lim_Kang_Xiong_Niyato_Yang_Shen_Miao_2022}. This has highlighted the need for resource allocation by MSPs. A key concern in resource allocation is supporting synchronization between the Metaverse and the real world. For instance, in~\cite{Han_Niyato_Leung_Miao_Kim_2022}, the digital twin association was discussed as an important factor affecting immersion and users' quality of experience (QoE). Similarly, the importance of dynamic resource allocation frameworks for Metaverse services was emphasized in~\cite{Du_Liu_Niyato_Kang_Xiong_Zhang_Kim_2023}. These works demonstrate the significance of resource allocation and synchronization for optimal user experience in Metaverse services.
\subsection{3D human pose estimation }
HPE is a task in computer vision that involves identifying the position and joints of a human body in a given scene. Various approaches have been proposed, such as the top-down approach~\cite{chen20222d}, which recognizes the human as an object first, then identifies the joints and forms the skeleton. Alternatively, the bottom-up approach~\cite{chen20222d} recognizes the human joints first and assembles them into the skeleton. In 3D pose estimation, one approach involves using stereoscopic or multiple cameras to obtain depth information to establish the 3D skeleton. Another approach is to use a monocular camera to obtain a 2D skeleton and perform 2D-3D lifting to reconstruct the 3D pose~\cite{Tome_2017_CVPR, 8954163}. These methods have been applied in various applications, such as fitness and rehabilitation, and Metaverse services. The emergence of these skeleton capture techniques has created ample opportunities for vision-based avatar associations.

\subsection{Contest theory}

Contests can be defined as games in which participants invest resources with a certain probability of winning prizes. This concept has been widely utilized for solving different problems. For instance, authors in~\cite{8734050} utilized contest theory to balance requests and services among users in a service exchange application with low-demand service. Moreover, a contest-based incentive mechanism was introduced in~\cite{10007890} to maximize participants' efforts in improving the QoS of MSPs.

Inspired by the above studies, this paper proposes a novel incentive framework in a Metaverse service by using a joint approach of the contest theory and deep reinforcement learning. Our approach utilizes the semantic encoding method through pose estimation to minimize the quantity of sensing data that requires transmission. Additionally, we present a dynamic resource allocation strategy based on contest theory for MSP avatar rendering service to further enhance the QoS.
\begin{figure}[t!]
\centerline{\includegraphics[width=0.4\textwidth]{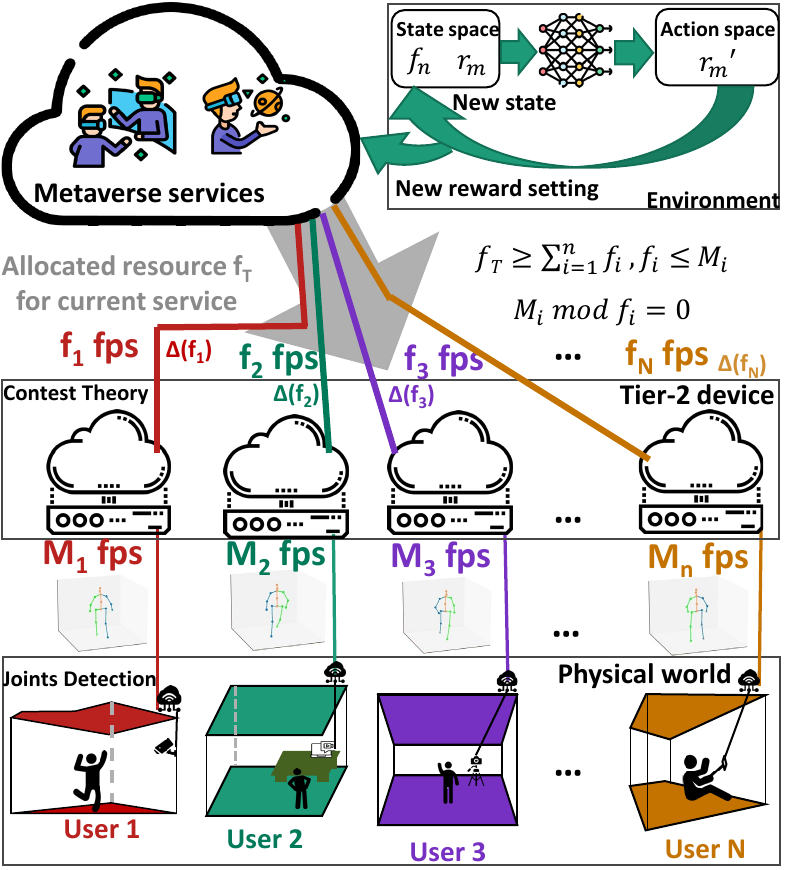}}
\caption{System diagram of the proposed framework }
\label{systemdigram}
\end{figure}
\section{SYSTEM OVERVIEW}

\subsection{Cameras in Physical World}
As illustrated in Fig.~\ref{systemdigram}, 3D body key-points detection Cameras from Closed Circuit Television (CCTV), computers, and smart devices capture real-time images to obtain visual data. Such data will then be encoded by extracting users' pose skeletons to obtain semantic data. The extraction is performed by smart devices or MSPs :
\begin{itemize}
\item The extractions  are performed by sensing devices: Capturing images, processing to extract the user's body keypoints, and transmitting the extracted semantic pose keypoints will experience large latency due to limited storage and computing resources.
\item The extractions are performed by the MSPs: The captured images instead of encoded semantic information is transmitted to MSPs. This will create a large data flow and threat to users' information privacy since visual images contain lots more information than the needed keypoints.
\end{itemize}

\subsection{3D Body Key-points Detection Encoding at Tier-2 Devices}
\begin{figure}[h]
\centerline{\includegraphics[width=0.45\textwidth]{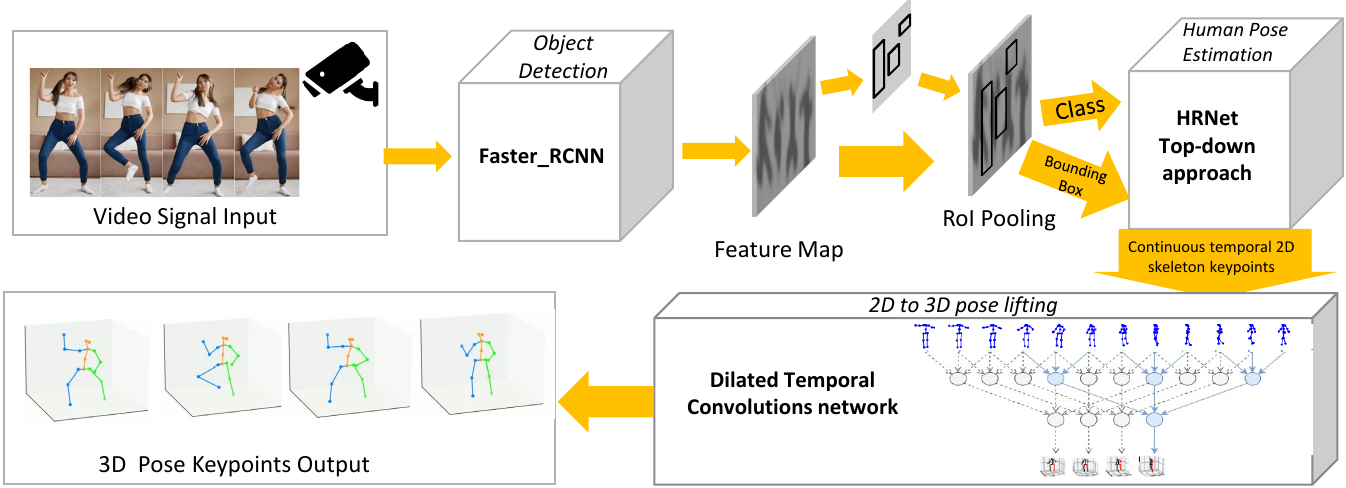 }}
\caption{Structures of 3 stages encoder }
\label{encoder}
\end{figure}

One of this paper's main focuses is capturing and maintaining users' motion characteristics for associating digital twin avatars. As illustrated in Fig.~\ref{encoder}, to extract such semantic pose data from the received visual data, Tier-2 Devices such as edge nodes or fog nodes conduct 3D Body Key-points recognition with respect to the transmitter. Consider frequently only a monocular camera is available, a monocular 3D Body Key-points extraction encoder proposed by Meta (formerly Facebook)~\cite{8954163} is deployed. This semantic encoder of the system consists of three stages: a human bounding box detection model, a top-down HPE model, and a 2D-3D keypoints lifting model. These models \cite{mmpose2020} are used as encoders before the incentive mechanism, and the selection of the models depends on the available computational power and specifications of the tier-2 devices. Other models may be substituted as needed to achieve the desired levels of accuracy and precision for the respective service.


\subsection{Contest Theory}

MSPs typically serve multiple services with multiple users simultaneously. The MSP's resources are distributed among users in a fair manner which means those users who move rapidly or have significant movements will experience nonconsecutive avatar updates. Meanwhile, users who move slowly or have smaller movements will receive extra resources to update their avatars. In the worst case, lots of resources are wasted on static motion users. Current solutions include increasing the smoothness of avatar linkage by allocating enough resources for all users assuming they have significant movement. However, such a practice could decrease the total number of users a single server could accommodate.

Note that keypoint upload rates directly impact avatar refresh rates and rendering of MSPs, which further affect the QoS. Thus, we introduce an incentive mechanism for tier-2 devices in the service framework based on contest theory. In order to win the award, the transmitter needs to upload extracted keypoints at the proper frequency. As a result, this mechanism improves MSP's overall rendering and refresh resource allocation, which further enhances QoS.

\section{Contest Theory-based resource Allocation}
To improve the user's experience with Metaverse services and optimize the resource usage of MSPs, an efficient incentive mechanism is designed to encourage tier-2 devices to upload semantic data at an appropriate rate. In particular, Our objective is to incentivize all contestants to choose a suitable update frequency for all users with a fixed total payment, rather than upload as frequently as possible. A suitable solution to this problem is to use payments to host a contest among transmitters. We define the contest as a game in which contestants, i.e., the tier-2 devices for an individual user, must choose a suitable uploading frequency to earn awards based on the down-sampling loss. This section examines the payoff of tier-2 devices and the payment of MSPs utilizing contest theory~\cite{contest}.
\subsection{Capability, effort and Award}
Under the aforementioned scenario, the contestants are tier-2 devices and respective exerted efforts are the semantic data updating frequency. Let $f_n$ denote the uploading frequency of the $n^{\rm{th}}$ contestants in $T$ updates. Therefore,  $n^{\rm{th}}$ contestant's cost function can be defined as a twice differentiable function, i.e., $C(a_n,f_{n})$ complying with~\cite{contest}
\begin{equation}\label{eq:1}
\left\{\begin{matrix}
\frac{\partial C(a_n,f_n))}{\partial f_n}> 0,\\
 \frac{\partial C(a_n,f_n))}{\partial a_n}< 0,\\
 \frac{\partial^2 C(a_n,f_n))}{\partial a_n\partial f_n}< 0,
\end{matrix}\right.
\end{equation}
where $a_n$ denote the capability of $n^{\rm{th}}$ contestants. The inequality of (\ref{eq:1}) shows that the more capable a contestant is, the more likely it will choose a larger $f_n$ as upload frequency.
\begin{figure}[h]
\centerline{\includegraphics[width=0.48\textwidth]{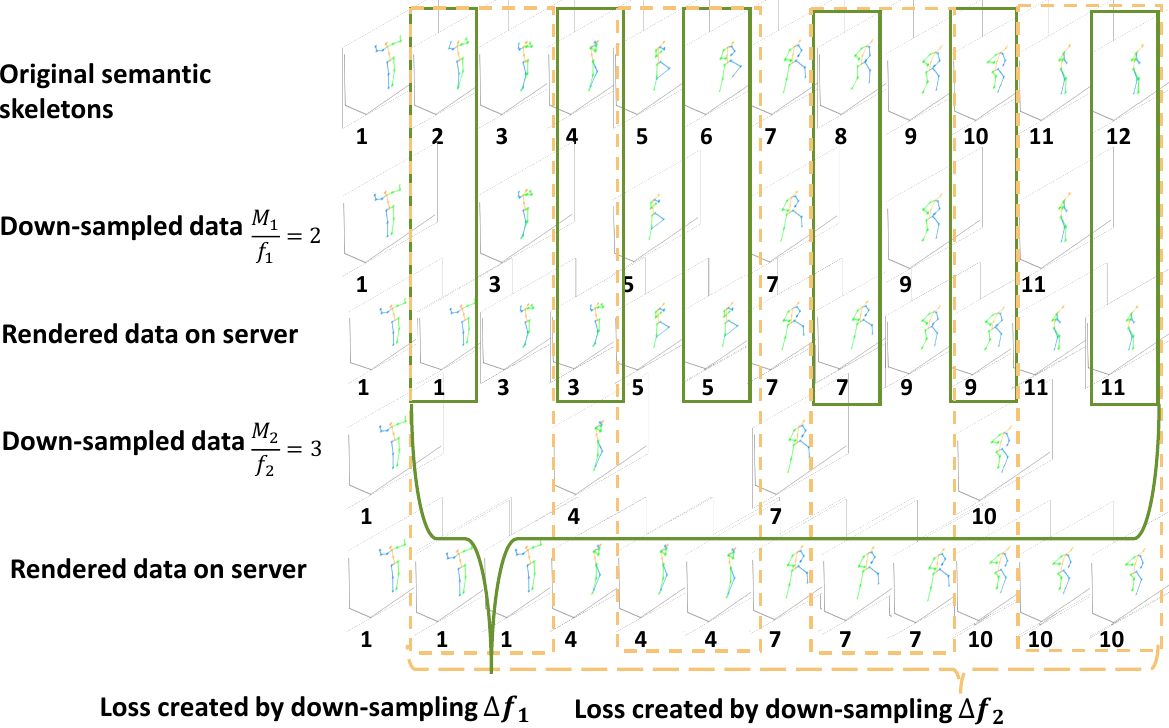}}
\caption{Two examples of how down-sampling loss is created }
\label{dsloss}
\end{figure}
Depending on the chosen effort $f_n$ that the tier-2 device Exerted, there is the loss between the real semantic signal and the sampled signal as illustrated in Fig.~\ref{dsloss}. Given the same down-sampling frequency, users with larger movements will create more loss compared to the viewed signal. Such loss reflects the movements of each user's motion and thus is regarded as the capability of each user. Let $(X_{ij},Y_{ij},Z_{ij})$ be the 3D coordinates of $j^{\rm{th}}$ keypoint out of $k$ keypoints of the $i^{\rm{th}}$ updates, and $(X'_{ij},Y'_{ij},Z'_{ij})$ as the rendered 3D coordinates according to $f_n$ on Metaverse server. Capability $a_n$ can be expressed as
\begin{equation}\label{eq:cap}
a_n=\Delta(f_n)=\sqrt{ \frac{1}{T}\times\sum_{i=1}^{T}\sum_{j=1}^{k}d},
\end{equation}
where $d=\left ( X_{ij}-{X}'_{ij} \right )^{2}+\left ( Y_{ij}-{Y}'_{ij}\right )^{2}+\left( Z_{ij}-{Z}'_{ij} \right )^{2}$, $\sqrt{d}$ is the euclidean distance between 3D coordinates between the original semantic skeleton keypoint and the rendered down-sampled skeleton keypoint.

As the capability is dependent on the chosen uploading frequency, for a given user with a fixed movement size, a more capable user will choose a larger frequency resulting in a lower loss. By taking (\ref{eq:1}) into consideration, the following conclusions can be drawn: the cost function can be expressed as $C(a_n, fn)=f_n/a_n$. Sorting $N_t$ contestants in descending order according to exerted efforts to obtain the effort list $\{f_1,f_2,\ldots,f_{N_T}\}$. The $n^{\rm{th}}$ contestant receive award $r_n(1\leq n \leq N_A \leq N_T) $, where $N_A$ is the number of available awards. As the reward is given in descending order, $r_1\geq r_2\geq\cdots\geq r_{N_A}$, and when $n>N_A$,$r_n=0$. The total award is fixed as it is often necessary for game theory and mechanism design to ensure that the total reward paid to participants is fixed. This rule is implemented to ensure a fair game is played and to prevent collusion between users~\cite{contest}. Thus the total award is expressed as $r_T=\sum_{n=1}^{N_A}r_n, r_T=R$.
\subsection{Utility}
The utility of the $n^{\rm{th}}$ contestant can be derived as follows:

 \begin{equation}
 \label{eq:utility}
U(a_{n},f_{n})= \begin{cases}
r_{m} -C(a_m,f_{m})), &{\rm for~ m^{\rm{th}}~rank~user,}\\
-C_{n}(f_{n}),& n\geq N_A.
\end{cases}
\end{equation}
\\
Under such a scenario, the users only possess knowledge of their own capability without access to others' data. The cumulative distribution of capability in the population is represented by a continuous function $\mathcal{P}(\Delta(f_n))$. Here, we assume that the $\mathcal{P} $ follows a uniform distribution~\cite{10007890}. The considered distribution is easily extensible to other distributions. For the $n^{\rm{th}}$ contestant, the probability that the other contestant's capability is smaller than its capability, i.e.,$\mathcal{P}(\Delta(f_n))$ is 

\begin{equation}
\label{eq:CDF:P}
{\mathcal{P}}({\Delta(f_n)}) = \begin{cases}
\frac{{max_\Delta(f_n)}-\Delta(f_n)}{max_{\Delta(f_n)}}, & 0\leq \Delta(f_n) \leq max_{\Delta(f_n)}, \\
0, & {\rm{otherwise}}.
\end{cases}
\end{equation}
With the obtained $\mathcal{P}(\Delta(f_n))$ and fixed payment pool, the expected number of contestants who win before the $n^{\rm{th}}$ contestant is $\left(N_T-1\right)\left(1-\mathcal{P}(\Delta(f_n))\right)$. Since the reward pool is fixed, the expected payment received by the $n^{\rm{th}}$ contestant is equal to the probability of winning $i^{\rm{\rm{th}}}$ payment($i-1$ numbers of contestants larger than $n^{\rm{th}}$ contestant) times the respective $i^{\rm{th}}$ payment. Therefore, the expected payment received by the $n^{\rm{th}}$ contestant can be calculated as follows:
 \begin{equation}
 \begin{aligned}\label{eq:expect}
 E(\Delta(f_n),r_m)=& \sum_{i=1}^{N_{A}}u(r_{m})\binom{N_{T}-1}{i-1}\\
                    &\times {\mathcal{P}}^{N_{T}-i}(\Delta(f_n))\\
                    &\times(1-\mathcal{P}(\Delta(f_n)))^{i-1}.
\end{aligned}
\end{equation}
\subsection{Optimal Loss Analysis}
As illustrated in Fig.~\ref{systemdigram}, the effort of contestants can be measured in terms of the semantic keypoints uploading frequency. Let $M_n$ denote the frequency that the $n^{\rm{th}}$ contestant receives the keypoints data and $f_n$ be the uploading frequency. Due to the sampling nature of visual information, $f_n$ must be a factor of $M_n$ to avoid uneven sampling. Furthermore, the total effort of all users cannot exceed the allocated resource $f_T$. With the constraints of average sampling of the visual signal, the available effort for selection $f_{n}$ is limited to the factor of $ M_{n} $. With $\mathbb{F}(M_{n})$ denoting the factor sets of $ M_{n} $, the effort is selected as follows:
 \begin{equation}
 \label{eq:choosef}
 \begin{aligned}
f_{n}=\arg\max E(\Delta(f^*),r_m), \quad f^{*} \in \mathbb{F}(M_{n}).
\end{aligned}
\end{equation}
Under the setting of the previous scenario, the server could easily adjust the resource allocation by adjusting the award setting $r_m$. The server adjusts the awards to minimize the sum of down-sampling-induced loss by solving
\begin{equation}
\label{eq:rmsetting}
\begin{aligned}
\min_{N_{A},r_{1},\ldots,r_{N_{A}}}~&\sum_{n=1}^{N_{T}}&\Delta(f_n,r_n) \\
\rm {s.t.}~&\sum_{n=1}^{N_T}&f_n \leq f_T ,~
&\sum_{m=1}^{N_A}&r_m = R.
\end{aligned}
\end{equation}
\subsection{Optimal award setting Analysis}

The analysis of the award settings poses a challenge as it is a mixed-integer linear programming problem, which traditional numerical tools may not be able to accommodate. Therefore, in this section, we use the DQN to solve (\ref{eq:rmsetting}) and obtain the optimal award setting that minimizes the sum of the Loss, even when dealing with a flexible range of users and services.

\textbf{State space:} The state space is a set of values that represent the current award setting $\left[r_1,r_2,\ldots ,r_{N_T}\right]$ and the effort exerted by the users $\left[f_1,f_2,\ldots,f_{N_T}\right]$.

\textbf{Action space:} The action space is supposed to be constructed by the award setting of the receiver: $\left[ r_1,r_2,\ldots,r_{N_T}\right]$. However, even with the limitation of fixed total reward $R$, the action is still too large to explore. In the case that $R=100$ and $N_T=4$, the number of possible actions is 103883550. Such a large action space will lead to difficulties in converging. Thus, we set the action to $\left[i_1,i_2,\ldots,i_{N_T}\right]$, where values of $i_n$ could be $1$ (Increase), $0$ (Stay), and $-1$ (Decrease). Such a design greatly reduce the action space. For example, when $N_T=4$, the number of possible actions is only 19.

\textbf{Reward function:} Rewards are set based on both the objective function and constraints related to (\ref{eq:rmsetting}). To be able to determine the optimal award setting between the instant reward term and the long-term reward term, the requirement-aware reward function is defined as

\begin{equation}\label{eq:reward}
reward(f_n,r_m)=\left\{\!\!\!\!\!\!\begin{matrix}&&0, &&{\rm{Condition 1}}\\
 &&c\times\frac{1}{\sum_{n=1}^{N_{T}}\Delta(f_n,r_m)},&&{\rm otherwise}
\end{matrix}\right. \\
\end{equation}
where c is constant, and Condition 1 is 
$\sum_{n=1}^{N_T}f_n < f_T$ or $\sum_{n=1}^{N_A}r_n > R$.

For clarity, the details of the proposed DQN-based approach are summarised in Algorithm 1. Additionally, we perform the analysis for the algorithm's time complexity~\cite{9465768}. We assume that the Deep Neural Network (DNN) deployed in DQN contains $L$ number of layers, $N_{l_{\rm{th}}}$ number of neurons per layer, and $\overline{I}$ number of local iterations each training. The training time complexity of such a Q-network is $O\left(\overline{I}\sum_{l=1}^{L}N_lN_{l+1}\right)$. The inference time complexity is $O\left(\overline{I}\right)$.
\begin{algorithm}[t]
\caption{The Proposed DQN-based Approach}\label{alg:DQN}
\textbf{Input:}  Current State $s=\left[  r_1,r_2,\ldots,r_{N_T},f_1,f_2,\ldots,f_{N_T}\right]$\\
\textbf{Output:} Action $a=\left[i_1,i_2,\ldots,i_{N_T}\right ]$
\begin{algorithmic}[1]
\While {episode = 1 to max episode} 
\State Observe initial state $s^t$, $t = 0$;
\For{step t = 1 to max time step}
\State Obtain updated award sets from action and previous state $\left[r_1+i_1,r_2+i_2,\ldots,r_{N_T}+i_{N_T}\right]$;
\State Take actions with updated award sets based on (\ref{eq:choosef});
\State Observe the reward $r^t$ according to (\ref{eq:reward});
\State Save $\left(s^t,a^t,r^t,s^{t+1}\right)$ into experience pool;
\State Update the new current state $s^t = s^{t+1}$;
\State Sample $X$ transitions from the experience pool;
\State Compute the expected $Q$ values
\State Update DNN parameters $\mathbf{\theta}$ via Optimiser;
\EndFor
\EndWhile
\end{algorithmic}
\end{algorithm}

\section{Experiment and results}
\subsection{Experiment setup, Scenario and Data collection}
\begin{table}[!ht]
 
\caption{Parameters For Experiment settings.}
\label{table:parameters}
    \centering
    \begin{tabular}{|c|c|c|c|c|}
    \hline
        \textbf{Parameters} & \multicolumn{4}{c|}{\textbf{Values}} \\ \hline
        \textbf{Users} & 1  & 2  & 3  & 4  \\ \hline
        \textbf{Video actions} &  Run   &  Dance   &  Wave   &  Stand   \\ \hline
        \textbf{Number of users$N_T$} & \multicolumn{4}{c|}{4} \\ \hline
        \textbf{Original frequency$M_n$} & \multicolumn{4}{c|}{60 fps} \\ \hline
        \textbf{Key points extracted$k$} & \multicolumn{4}{c|}{17}  \\ \hline
        \textbf{Total resource $f_T$} & \multicolumn{4}{c|}{120 fps}  \\ \hline
        \textbf{Total award$R$} & \multicolumn{4}{c|}{100}  \\ \hline
    \end{tabular}
\end{table}

Four 5-second videos with a resolution of $1080\rm{px}\times1908\rm{px}$ and 60 frames per second (fps) were collected using real-life cameras as input for our system. To extract each user's semantic skeleton information, we followed the three-stage process outlined in Fig.~\ref{encoder}, resulting in a total of 300 frames of skeleton information for each user. Each frame consisted of 17 3D coordinates, providing data for later analysis. Besides the semantic skeleton data, other important parameters used in the study are summarized in Table \ref{table:parameters}. Although rendering resources are not typically measured in fps, we adopt this metric to reflect the resources allocated to each service. Which is based on the observation that the processing and rendering resources required for each frame are consistent, given the constant size of each frame of skeleton information. Thus, we use the number of fps that the server can accommodate and render with the allocated resources as the measure.

\subsection{Performance Evaluation}
\begin{figure}[t!]
\centerline{\includegraphics[width=0.4\textwidth]{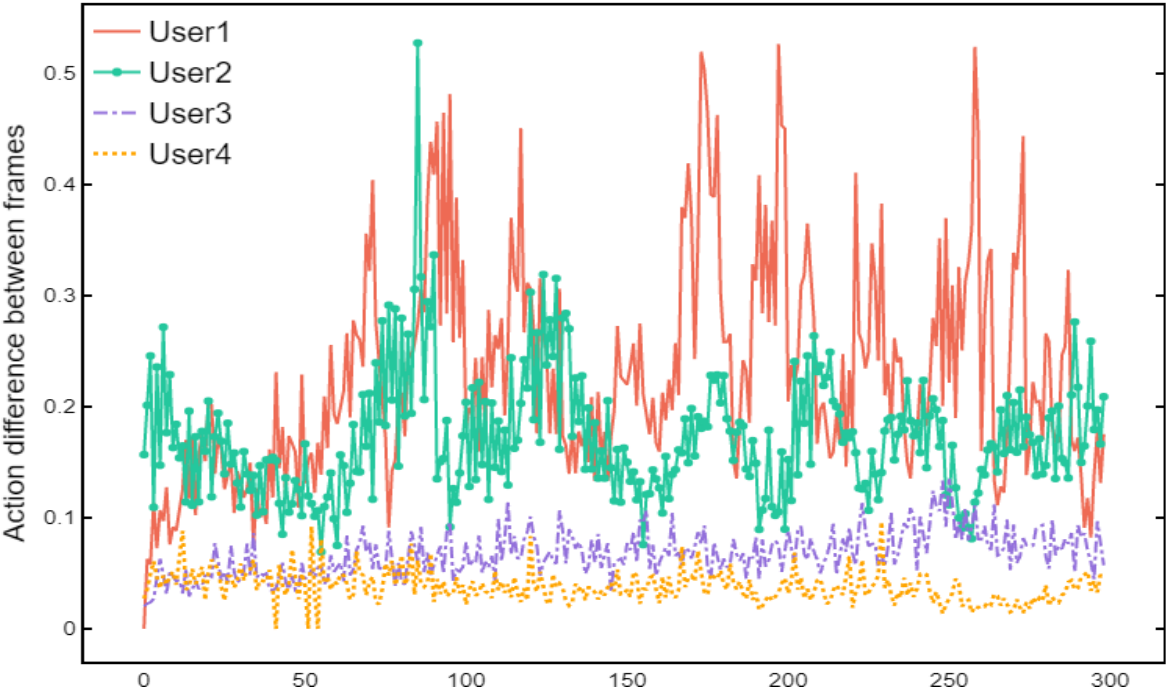}}
\caption{Comparison of motion difference among 4 users. }
\label{fig:usersdifference}
\end{figure}
\subsubsection{Users Movement}
The Euclidean distance between each frame's pose is illustrated in Fig.~\ref{fig:usersdifference} to validate the movements of each user. As shown in Fig.~\ref{fig:usersdifference}, User 1 possesses the largest average movement followed by users 2, 3 then 4. In a real-world scenario, factors such as the user's distance to the camera, body size, and camera angle can affect the parameters. To mitigate these disturbances, a 2D-3D lifting process is implemented, which eliminates the distortions by using an origin-rebased function with a Dilated Temporal Convolutions network~\cite{8954163}. This results in a skeleton with a unified height, which captures only the user's actions and eliminates the effect of the camera angle. This provides an added layer of privacy, as the height and camera position is not made known to MSPs.

\subsubsection{User chosen effort under different award setting}
\begin{figure}[t!]
  \centering
  \subfigure[Winner take all]{\includegraphics[width=0.13\textwidth ]{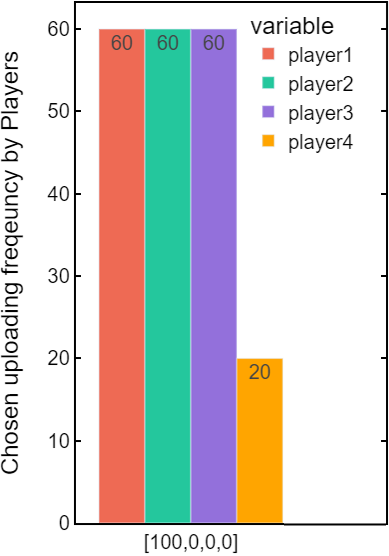}    \label{fig:userchosen1}}\quad
  \subfigure[Winner take half]{\includegraphics[width=0.13\textwidth ]{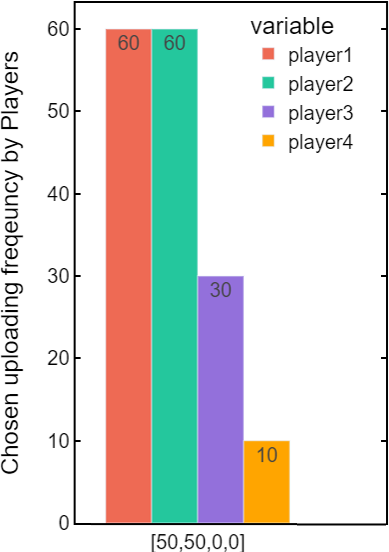}\label{fig:userchosen2}}\quad
    \subfigure[Average to all]{\includegraphics[width=0.13\textwidth ]{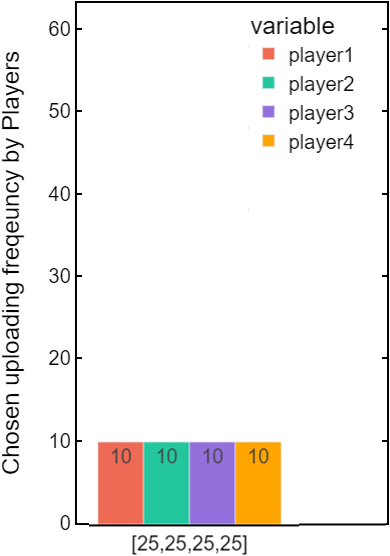}\label{fig:userchosen3}}
   \caption{Users chosen effort under the different schemes}
  \label{fig:chosenfn}
\end{figure}

We present experimental results on a system with varying reward settings, as specified in Table~\ref{table:parameters}. The results, presented in Fig.~\ref{fig:chosenfn}, show that when no constraints are imposed on the reward distribution for $f_T$, awarding the prize solely to the first-place winner leads to users putting in their best efforts according to their abilities. Conversely, when the prize is evenly distributed among all users, all users exert minimal effort irrespective of their capabilities, which aligns with the finding in~\cite{10007890}.

Put simply, when the reward is allocated only to the first winner $\left(\left[100,0,0,0\right]\right)$, all users try to upload as frequently as possible to minimize down-sampling loss and increase their chances of winning as shown in Fig. \ref{fig:userchosen1}. However, when the reward is evenly distributed among users $\left(\left[25,25,25,25\right]\right)$, each user receives the same reward regardless of their effort, resulting in all users choosing the lowest uploading frequency and exerting minimal effort as shown in Fig. \ref{fig:userchosen3}.

Furthermore, when the award is evenly distributed to the first two winners $\left(\left[50,50,0,0\right]\right)$, users with higher capability choose to upload more frequently to secure the award, while lower capability users upload less frequently to reduce the cost of uploading as shown in Fig. \ref{fig:userchosen2}. Such varied behaviours under different award setting scenarios demonstrate that users exert suitable effort based on the award settings.

\begin{figure}[t!]
  \centering
  \subfigure[Reward analysis]{\includegraphics[width=0.2\textwidth]{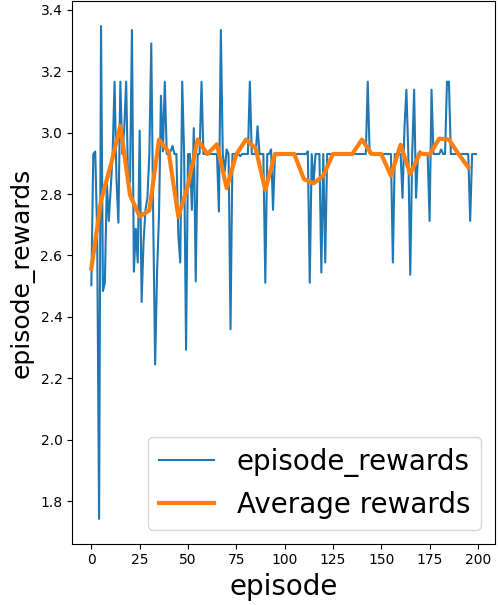}    \label{fig:dqnREWARD}}\quad
  \subfigure[Loss analysis]{\includegraphics[width=0.2\textwidth]{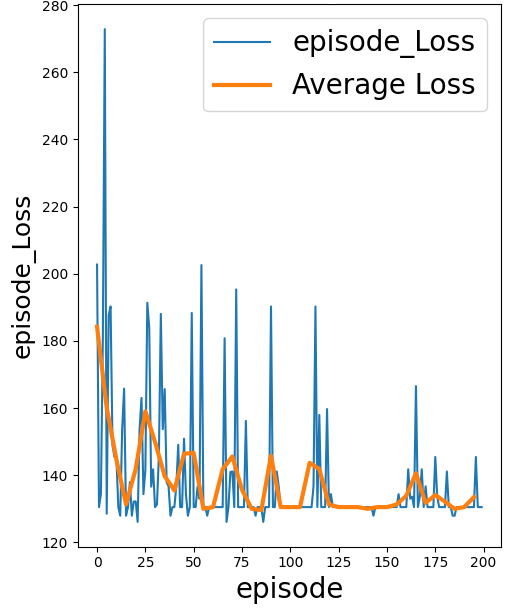}\label{fig:dqnLOSS}}
   \caption{Rewards and down-sampling losses over episodes}
  \label{fig:rewardandloss}
\end{figure}

\begin{figure}[t!]
\centerline{\includegraphics[width=0.4\textwidth]{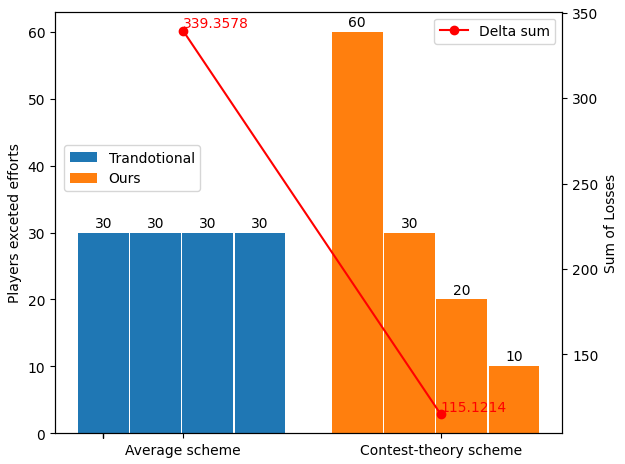}}
\caption{Comparison between traditional average distributed resource and Contest theory based approach}
\label{fig:dqnresults}
\end{figure}

\subsubsection{Comparison of the amount of data before and after pose estimation}
With the proposed semantic encoding approach, a significant reduction in the data amount can be achieved while preserving system performance. A single frame image signal, which initially consists of approximately 8.243 megabytes of data ($1080\times1908\times32/8$), is now reduced to 51 bytes ($17\times3\times1$) of encoded data. Additionally, such encoding mechanism ensures the security of information by not allowing any MPSs to obtain excessive visual information from users. Video information no longer has to propagate through the entire network, thereby reducing the threat to confidentiality.

\subsubsection{DQN optimization on reward setting}

DQN is effectively utilized in the optimization process to determine the optimal reward setting that incentivizes the desired level of effort. As illustrated in Fig.~\ref{fig:dqnREWARD}, the results show a clear improvement in the average rewards obtained per episode, which increased steadily and converged over time. Additionally, as depicted in Fig.~\ref{fig:dqnLOSS}, the average down-sampling loss decreased consistently and eventually converged. The comparison between the averagely distributed resources and the user-chosen effort under DQN optimized reward setting, as shown in Fig.~\ref{fig:dqnresults}, highlights the significant improvement achieved, with the sum of losses decreasing from $339.3578$ to $115.1214$. With a significant reduction of 66.076\% in the total loss achieved through dynamic resource allocation, the effectiveness of utilizing DQN for optimizing reward settings is demonstrated. The results also highlight the ability of DQN to effectively reduce fluctuations in the discrete action space, leading to convergence over time.

\section{Conclusion}

In this paper, we propose a novel framework that leverages a joint approach of the contest theory and deep reinforcement learning to incentivize user effort in a Metaverse service. The proposed framework offers the advantage of reduced data transmission, increased privacy for users, and optimized reward settings that can effectively incentivize users to provide semantic data to the MSPs. The experiment results have shown the efficacy of the semantic encoding approach in reducing the amount of transmitted data while maintaining system performance. The results also highlight the significant impact of reward distribution on users' efforts, emphasizing the importance of reward settings in motivating users. Moreover, the DQN optimization has demonstrated the capability of optimizing reward settings to encourage the desired level of user effort, resulting in a substantial reduction in the sum of losses. These findings suggest the potential of the proposed approach for real-world applications.
\bibliographystyle{IEEEtran}
\bibliography{citations}
\end{document}